A search for the Mpemba effect: When hot water freezes faster then cold water

James D. Brownridge
*Department of Physics, Applied Physics, and Astronomy, State University of New York at Binghamton, P.O. Box 6000 Binghamton, New York 13902-6000, USA*

Abstract:

An explanation for why hot water will sometime freeze more rapidly than cold water is offered. Two specimens of water from the same source will often have different spontaneous freezing temperatures; that is, the temperature at which freezing begins. When both specimens supercool and the spontaneous freezing temperature of the hot water is higher than that of the cold water, then the hot water will usually freeze first, *if all other conditions are equal and remain so during cooling*. The probability that the hot water will freeze first if it has the higher spontaneous freezing temperature will be larger for a larger difference in spontaneous freezing temperature. Heating the water may lower, raise or not change the spontaneous freezing temperature. The keys to observing hot water freezing before cold water are supercooling the water and having a significant difference in the spontaneous freezing temperature of the two water specimens. We observed hot water freezing before cold water 28 times in 28 attempts under the conditions described here.

## I. Introduction

Two identical containers holding the same amount of water at different temperatures are placed in a freezer, and you determine that the hot water freezes first. Did you observe the Mpemba effect?[1] The Mpemba effect describes the phenomenon of a sample of hot water freezing faster than a specimen of cold water. To understand why we ask this question, we encourage the reader to read the Monwhea Jeng review paper, **Hot water can freeze faster than cold,** for background and perspective.[2] In this paper; we will attempt to explain why hot water can freeze faster than cold water when all other conditions are identical. We will also demonstrate, by showing experimental results, why, for over 2000 years, it has been so difficult to reproduce this result in the same laboratory, let alone at different laboratories and with different investigators.

Before starting the search for the Mpemba effect, the meaning of freezing faster or freezing first must be addressed. Does it mean the onset of the release of the latent heat of freezing, or the appearance of the first ice crystal or the solidification of the last bit of liquid water? If you selected one of the above as your answer, how would you observe the time at which that event occurred in a specimen of water? For example, in Fig. 3, how long did it take for the water to freeze? Now, imagine that we have combined the results from eight different laboratories in which everything was **identical, except** for the positions of the thermocouples in the water. In this case, each laboratory would report a different time of freezing. These differences were a major barrier to obtaining reproducible results during the early phase of this study. A difference in position of only a few millimeters can be significant depending on the size of the container.

Because, of the uncertainty in what is meant by "freezing first" or "freezing faster," we will define the time of freezing as the time when the latent heat of freezing is released. The cooling curves in this paper clearly show when the latent heat of freezing is released. The exact time when the first ice crystal appears or when the last few molecules of liquid water become solid ice is not always apparent. However, a sudden rise in temperature from several degrees below $0^o$ C indicates that the latent heat of freezing was released as the specimen supercooled, as seen in Fig. 1. In the rare case when a specimen does not supercool, the temperature suddenly stops decreasing at $0^o$ C, indicating that the sample is freezing.

With the tip of a thermocouple in a specimen of water, the time that the latent heat of freezing is released can be determined to within 100 ms as the water supercools a few $^o$ C, as seen in Fig. 5.

The fact that hot water will sometimes freeze before cold water has been observed and debated since around 350 BC; see reference 2 and the references therein. The question of why hot water can freeze before cold water when all other conditions are as equal as possible is still debated in both popular and scientific literature.[1-5] Several good discussions of the history of the Mpemba effect and why there is not a generally accepted answer can be found in reference 2 and the references therein.

Several years ago, a colleague asked me if it was true that hot water could freeze before cold water. I answered yes, and he asked why; to make a long story short, I did not know. However, I assured him that I would have the answer soon. That was more than ten years ago. Now, after many years of laboratory work freezing hot and cold water, I am prepared to explain why hot water will sometimes freeze before cold water when one of two seemingly identical specimens are randomly selected for heating and both are placed in a cooling environment to freeze.

As it turned out, data suggesting an answer to this question was published by N. Ernest Dorsey in 1948.[6] Although he did not specifically address or answer the question, the answer is clear in his data. My laboratory results have confirmed several of his findings. Key among them: "…For each specimen there are one or more preferred temperatures at which spontaneous freezing occurs…" and "…Preheating the melt produces no certain effect upon it …"[6] In other words, if a specimen of water in a sealed container is heated and then cooled again, it may freeze at a higher, lower or the same temperature as it would have had it not been heated. Heating a specimen of water may or may not cause it to freeze before a cold specimen. Scientific and popular literature, as well as anecdotal reports, confirms this.[1-9]

Dorsey referred to "…preferred temperature at which spontaneous freezing occurs" as *motes* in 1948; they are now called nucleation sites.[6] He purposed that there may be one or more *motes* in a specimen of water and that they each may have a different spontaneous freezing temperature, which can be changed.

If each nucleation site (*mote*) in a specimen of water has a unique spontaneous freezing temperature[6], then a specimen of water should freeze when its temperature reaches the highest spontaneous freezing temperature among the nucleation sites, which is what we observed.

A quote from the review paper by Monwhea Jeng[2] explains why there is not a generally accepted mechanism for the Mpemba effect.

"Because there are so many factors that can be varied and the results of the experiments can depend sensitively on any of these factors, experimental results are varied and difficult to organize into a consistent picture..." Monwhea Jeng.[2]

Some of the factors are:

The definition of freezing first; is it the first ice crystal or is it when all liquid is solid?
The definition of the freezing time.
The position of the temperature-measuring device in the water.
The container shape.
The type of container.
Are all conditions except temperature truly identical, and do they remain so during cooling?
The loss of water by evaporation.
Does the hot container cause a change in the cooling rate?

In this research, we independently addressed each of these variables and more.

**II. The paradox**

Two "identical" containers holding the same amount of water by weight, one hot and one cold are placed in a cold environment to freeze. The hot water freezes first. Some time later, two "identical" containers holding equal amounts of water by weight, one hot and one cool, are placed in a different cold environment to freeze. The results of these two experiments are presented in Figs. 1 and 2. This ambiguity in which sample freezes first is typical of published experimental investigations of the Mpemba effect. The aim of this work is to identify the conditions under which hot water will consistently freeze before cold water. What follows is a description of a set of experiments that led to the identification of these conditions.

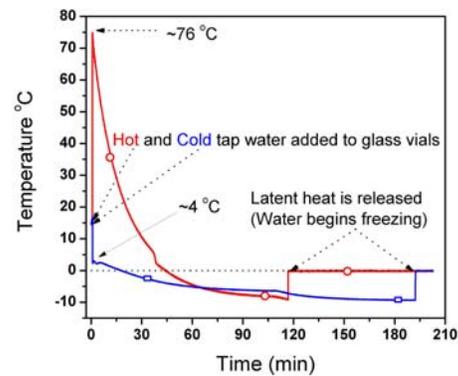

**Fig. 1.** Cooling curves of hot water and cold water. The same amount of water was added to two glass vials of equal weight. The vials were then suspended inside a freezer and allowed to air cool. The temperature of the air inside the freezer was -15°C.

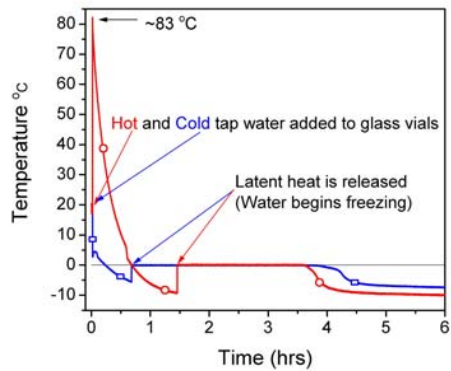

**Fig. 2.** Cooling curves of hot water and cold water. The same amount of water was added to two glass vials of equal weight. The vials were then suspended inside a freezer and allowed to air cool. The temperature of the air inside the freezer was -15º C.

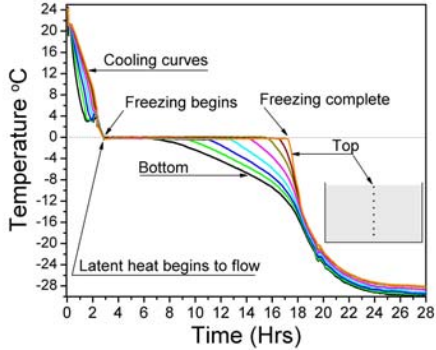

**Fig. 3.** The signal that latent heat has begun to flow is independent of the position of the thermocouple tip in the water (inset (a)). The shape of the cooling curves above freezing and the time to solid ice are, however, very sensitive to the thermocouple position. To generate these curves, eight thermocouples were placed in a specimen of water, each separated by 1cm in depth as depicted in the inset.

## III. Data Acquisition Systems

The data acquisition consisted of one Omega OMB-DAQ-3000 Series 1-MHz, 16-bit USB Data Acquisition module, two National Instruments PIC-6034E interface cards configured to accept eight Type K thermocouples each, two PASCO Xplorer GLX with eight temperature probes, two Keithley Instrument Mod. 155 Null Detector micro voltmeters, three freezers and a Lauda RM6-RMS Brinkmann Refrigerating Circulating Bath. Temperature and time data were recorded by computer, usually at a rate of 1 Hz, but more slowly for cooling rates less than 1º C/min. With these setups, we collected data for several thousand freeze/thaw cycles, as it took this many cycles to discover all of the things that could and did go wrong and to design a set of experiments that met our objectives.

## IV Experiment 1 Definition of "freezing first"

After settling on a definition of "freezing first", we needed an objective way to determine exactly when a specimen of water was frozen. Because we defined "freezing" as the instant that the latent heat of freezing begins to be released, the experiment was designed to objectively determine this time. In Fig. 4, we show a schematic diagram of the apparatus designed to measure the onset of the release of the latent heat of freezing in two simultaneous ways.

If a specimen of water supercools, we see a sudden rise in temperature toward 0º C. If the water does not supercool, then the temperature stops decreasing at 0º C. In either case, with the apparatus shown in Fig. 4, an emf voltage signal gets produced at the onset of the release of the latent heat of freezing; see also Fig. 5. Notice that this signal is observed three seconds before the temperature probe registers a change in temperature. The deeper the water supercools, the closer the voltage signal and the signal from the temperature probes occur in time, regardless of the size of the container. In a large container (>25 ml), a temperature probe may not signal that freezing has begun for some time. However, the electrical signal is transmitted at the instant the first ice crystal forms, no matter where it is located in the container.

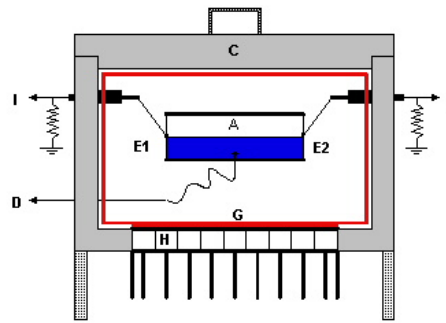

**Fig. 4.** Setup to measure the emf voltage produced the when latent heat of freezing is released. (A) Glass tube and water, (B) 500kΩ, (C) vacuum chamber, (D) thermocouple, (E) gold end caps, (G) copper box, (H) thermoelectric heater/cooler, (I) to voltmeters and computer.

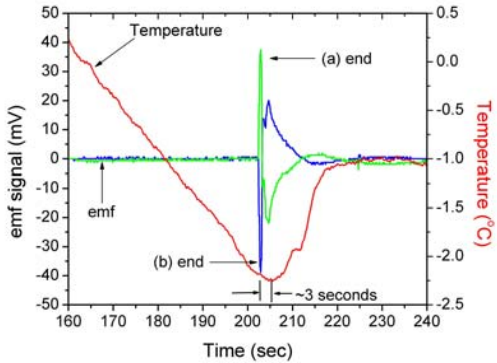

**Fig. 5.** The blue and green traces are emf signals vs. time as water cools. Notice the sharp spike in the emf signal just before the temperature began rising. The emf spike was produced at the onset of the phase change from liquid to solid and the release of the latent heat of freezing. The red trace is water temperature versus time.

With this emf signal, we can now objectively say when the water began freezing. In small containers, there is good correlation between the electrical signal and a thermocouple's response to the release of the latent heat of freezing. Also, the thermocouple's response time is short relative to the cooling time. Experience has shown that a still water specimen almost always supercools to less than -1° C, whereas if the water is stirred while cooling, it will freeze as soon as the temperature reaches 0° C. Therefore, since this work is with still water, we will rely on a thermocouple's response to the release of the latent heat of freezing to determine when a specimen of water freezes.

## V  Experiment 2   Time to freezing

Experiment number 2 was designed to determine whether the initial temperature of a specimen of clean water affects the time it takes to freeze. Clean water is defined as distilled, deionized water.

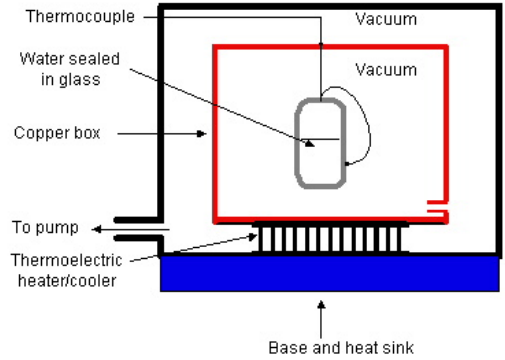

**Fig. 6.**   Experiment 2 setup

One milliliter of clean water was flame-sealed in a small Pyrex test tube, hereafter referred to as a vial. A type K thermocouple was affixed with epoxy to the outside of the vial below the water line. The vial was suspended in the center of a copper box that was affixed to a thermoelectric heater/cooler inside a vacuum chamber, pumped down to less than 1 mTorr. The idea was to heat and cool the vial and water without contact or physical disturbance. The vial was heated and cooled many times over a period of several weeks. The vial was supported by the thermocouple wires, and was heated and cooled by radiation. The side of the thermoelectric heater/cooler in contact with the copper box was heated or cooled depending on the polarity of the applied voltage.

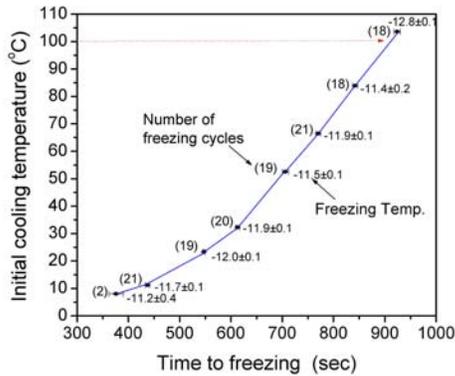

**Fig. 7.** Results of Experiment 2. Initial temperature of deionized, distilled water vs. time to freezing. The temperature and time were recorded once every second until the latent heat of freezing was released.

The data presented in Fig. 7 was collected from a single specimen of water that was randomly heated to temperatures between ~8° C and ~103° C and then cooled until it froze. This cycle was performed 138 times without disturbing the vial. The mean spontaneous freezing temperature was -11.80 ± 0.17° C, with a maximum freezing temperature of -11.2 ± 0.4° C and a minimum of -12.8 ± 0.1° C. The time to freezing was not reduced by heating this specimen of water; therefore, **no** Mpemba effect was observed. In fact, the hotter the water, the longer it took to cool to 0° C and ultimately freeze from a supercool state. This experiment was repeated in a completely different setup, using tap water, and still **no** Mpemba effect was observed; see Fig. 8.

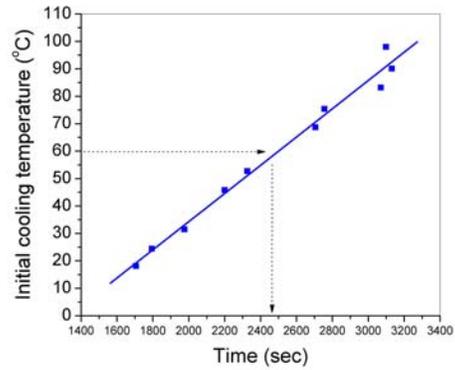

**Fig. 8.** Initial temperature of hard tap water vs. time to freezing.

The shapes of the curves in Figs 7 and 8 are quite different; I suspect this difference is due to the difference in cooling rates. However, these data show that each specimen of water had a narrow range of nucleation temperatures. The nucleation temperatures of the specimen used to produce the data presented in Fig. 7 were centered at -11.8° C. No attempt was made to determine if the errors associated with these measurements were due to different nucleation sites with different nucleation temperatures in the specimen or to the experimental setup. It appears that the specimen used to produce the data presented in Fig. 7 has a preferred spontaneous freezing temperature at about -11.80° C that was unaffected by heating to above 100° C, which was done 18 times in the 138 freeze/thaw cycles. Fig. 8 shows an independent confirmation that just heating water does not cause it to freeze faster than it would have had it not been heated.

With the number of variables minimized, **no** Mpemba effect was observed in a specimen of water sealed in a glass vial.

**VI. Experiment 3**   Hot water freezes first

So, why does hot water sometimes freeze before cooler water? In some cases, it is because of higher thermal conductivity.

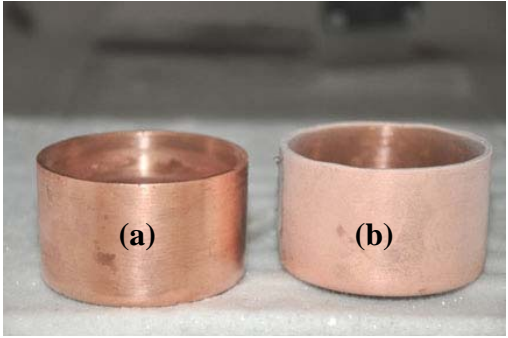

**Fig. 9.** Two copper cups, each weighing 125 g and containing 50 g of cold (a) or hot (b) water were placed inside a freezer at the same time. There was about 5 mm of frost on the floor of the freezer. Notice the melted frost under (b) and the frost on the cup; it is colder than (a)**.**

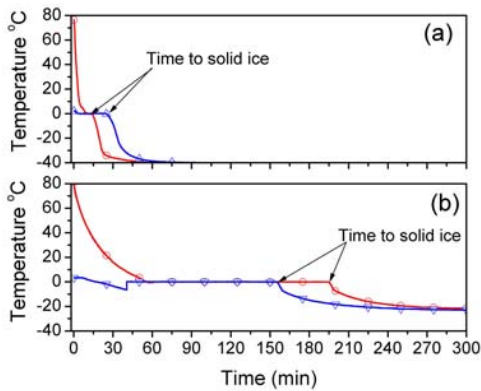

**Fig. 10.** Here we show the dramatic difference in the cooling rate when (a) a cup of hot hard water and a cup of cold hard water are placed in a freezer on a bed of freezer frost, vs. (b) when the same two cups are placed on an insulator and then put into the freezer on the same bed of freezer frost.

A typical example is presented in Fig. 10 (a). In Fig. 9, we show two identical coppers cups holding the same amount of water by weight. They are placed in a freezer on a bed of freezer frost. The hot cup causes the frost under the cup to melt, forming a pool of liquid water that soon freezes. The cooling conditions have just changed because the thermal conductivity of water is higher than ice and that of ice is higher than freezer frost. Now the hot-water container is in better thermal contact with the freezer floor than the cold-water container, and heat will now flow from it to the freezer faster. This may very well result in the hot water freezing first, as shown in Fig. 10(a).

Although this experiment began with identical conditions, with the exception of the initial water temperature, the hot water caused the conditions to change. The melting of the freezer frost produced a change in the thermal conductivity. If we prevent the hot-water container from melting more freezer frost than the cold-water container, then we should be able to maintain nearly identical conditions for the two containers. In Fig. 10(b), we show the results when the two containers were placed on an insulator, rather than directly on the frost, and observe the cold water freezing first. Water may not always freeze when its temperature falls to $0^o$ C. The temperature can fall below $0^o$ C before the onset of freezing; this phenomenon is known as supercooling. An example is shown in Fig. 10(b), in the cold water trace. Fig. 10(a) shows inconclusive results because the changing conditions of the experiment may have had a larger effect than the temperature difference.

**VII.   Experiment 4**  Cooling to $0^oC$

Will heating water cause it to cool to $0^o$ C more rapidly than cold water in the same cooling environment? *Newton's Law of Cooling forbids this effect.*

For the following measurement, equal amounts of hot or cold water were added to one or more of four specimen holders. The four specimens were held in separate compartments of one container. The container weighed 400 gm, far exceeding the 48 gm of water it held. The container was made of oxygen-free, highly thermally-conductive copper. Four shallow holes were milled into the solid copper, as shown in Fig. 11. Type K thermocouples were positioned in the center of each hole to record the temperature near the surface of the water. A copper lid covered the water to prevent ice crystals from falling into the water and initiating spontaneous freezing at $0^o$ C. In each freeze cycle, fresh water was added to the four holes and the copper block was placed in a freezer. The process was repeated many times with fresh specimens for several weeks, with three to five freeze cycles per day.

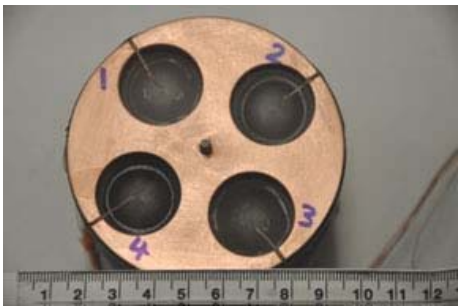

**Fig. 11.**     Apparatus used to compare the cooling rates of water samples at different initial temperatures.

In Fig. 12, we show results from simultaneously placing two hot and cold water specimens in a cooling environment. Notice that the temperature of the cold water initially rises and the temperature of the hot water falls. Once all four specimens reach the same temperature, they cool to $0^o$ C at the same rate. The time to cool to $0^o$ C is the same for hot and cold water when all other conditions are the same. Heating water does not cause it to cool faster than cold water; once the specimen reaches thermal equilibrium, the cooling rates are identical down to $0^o$ C.

In this case, the two cool-water specimens "froze first", which happened about 50% of the time when fresh specimens were used for each freeze cycle. This will be addressed later. When water cools to below $0^o$ C, the release of latent heat changes the cooling rate, as clearly shown in Fig. 12.

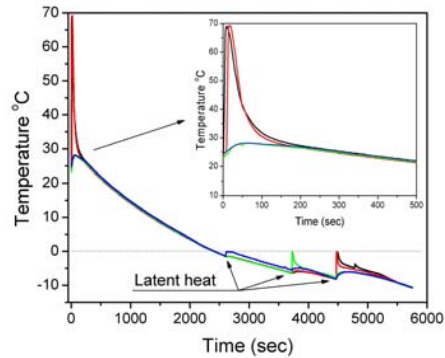

**Fig. 12.**     All conditions are identical, except for the temperature, for four specimens of tap water.
The inset shows the first 500 seconds of cooling before thermal equilibrium is reached at about 200 seconds. Thereafter, four specimens cooled at one rate until the temperature fell below $0^o$ C. The bumps in the curves below $0^o$ C are the releases of latent heat, the times at which each specimen froze. The thermocouples are about 1 mm away from the walls of the holders, as shown in Fig. 11.

**VIII.   Experiment 5**        Solutes, minerals, gaseous and sugared milk

If all conditions except one (temperature) are equal, then hot water cannot cool to $0^o$ C before cooler water. That would violate the laws of thermodynamics. However, if heating water changes the properties of the water or the conditions of the experiment then the cooling rate may be affected and then hot water may indeed cool to 0° C first.

To test this hypothesis, we placed four specimens of hard well water into the cooling chambers depicted in Fig. 11. Two of the samples were boiled until their original volumes were reduced by ~50%, and the other two were left unheated. The cooling curves are shown in Fig. 13; there was no difference in the cooling rates from the time that thermal equilibrium was achieved until the samples reached 0° C. We do see a change below 0° C, but it cannot be correlated with the initial temperature. To confirm that it is possible to see a change in the time it takes identical volumes of water to cool to 0° C, we added 0.3 ml of heavy water ($D_2O$) to each of two 12-ml specimens of deionized, distilled water, one of which was mixed by stirring. Unmixed $D_2O$ in $H_2O$ has been shown to reduce heat flow from $H_2O$[10]. The results are shown in Fig. 14, the red and green traces. Here, unmixed heavy water in the water specimen delays the transfer of heat from the water to the copper container. Once the heavy water is mixed by stirring, it has no observable effect. The cooling time of water with high concentrations of minerals such as calcium bicarbonate (hard water) is not affected by heating the water; see Fig. 13. We added the heavy water (Fig. 14) to confirm that our equipment was sensitive to a subtle difference in the water properties (stirred vs. unstirred). Milk-sugar mixtures, the essence of ice cream, cool to 0° C at a slower rate than clean water; see Fig.15. Whether the milk-sugar mixture freezes solid before the water depends on the temperature to which each specimen can supercool. In this case, the water froze first because it did not supercool as low as the milk-sugar mixture. In other cases, the milk-sugar mixture froze first.

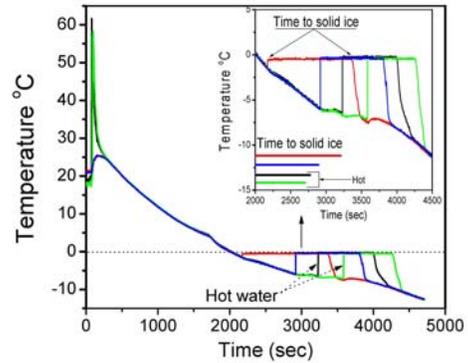

**Fig. 13.** Four well water specimens (all hard water) from the same source. Two were boiled, and two were never heated. The boiled water was degassed by boiling and transferred to the specimen holder while still hot.

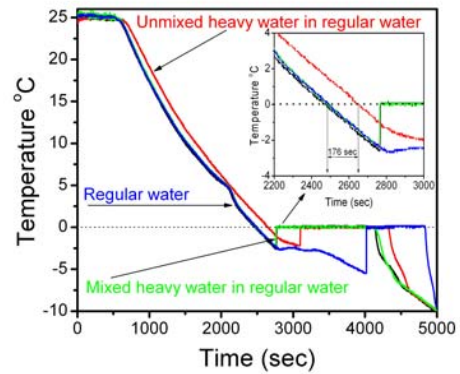

**Fig. 14.** Unmixed $D_2O$ in regular water (red) has the effect of increasing the time it takes water to cool to 0° C. Mixing the $D_2O$ in regular water (green) has no effect. The green trace is very hard well water, and the blue trace in deionized, distilled water.

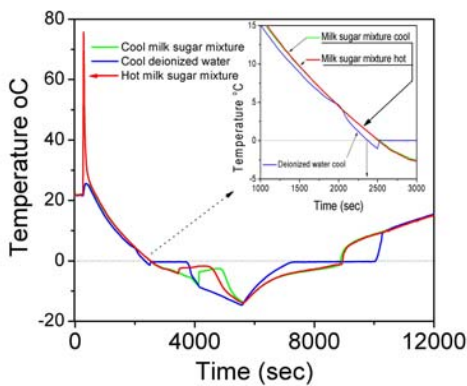

**Fig. 15.** Cooling curves of hot and cool milk-sugar mixture and water specimens cooled simultaneously in the copper holder shown in Fig. 11.

So, why does hot water sometimes freeze before cooler water even when all conditions are as equal as is practical?

**IX. Experiment 6**  Water in sealed containers

Five ml specimens of water were flame-sealed in six Pyrex test tubes. Three contained tap water, and three contained deionized, distilled water. The intent was to reduce the number of variables and to cool six vials at the same time in the same environment. It is assumed that no two vials will be the same, no matter how much care is taken. They were placed in the freezer as a unit to air cool. They were not in contact with each other; air was the cooling medium. The holder and vials were placed into and removed from the freezer as a unit. Care was taken not to shake or otherwise agitate the system, except when that was the object of an experiment. The freeze/thaw cycle was repeated several times over many weeks. With this arrangement, the vials could be shaken, heated or turned upside down in attempts to affect the spontaneous freezing temperature.

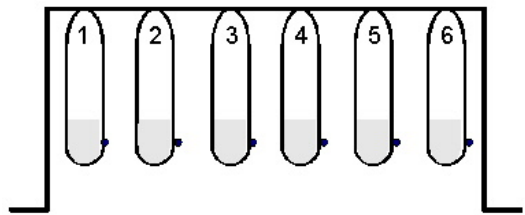

**Fig. 16.** Setup for simultaneously air-cooling six vials. The black dots indicate the locations of the thermocouples. Vials 1, 3 and 5 contain tap water, while 2, 4 and 6 contain deionized, distilled water.

The first objective was to determine the spontaneous freezing temperature for the water in each vial. This was accomplished by recording the temperature of each vial once per second until the water in all six vials froze. The spontaneous freezing temperature was then extracted from the cooling curves as the lowest temperature recorded before the temperature of the water began rising. See Fig. 21 for typical examples. The vials were cycled through freeze/thaw cycles 27 times in order to determine the spontaneous freezing temperatures. The results from each vial are presented in Fig. 17(a). The second objective was to see if vigorous shaking would affect the spontaneous freezing temperature. These results are presented in Fig. 17(b) and show that no major changes occurred in these specimens after 5 min of vigorously splashing the water from end to end inside the glass tubes. In other experiments, we have observed major changes produced by shaking different specimens in different containers. We will present data that show what we consider major changes later.

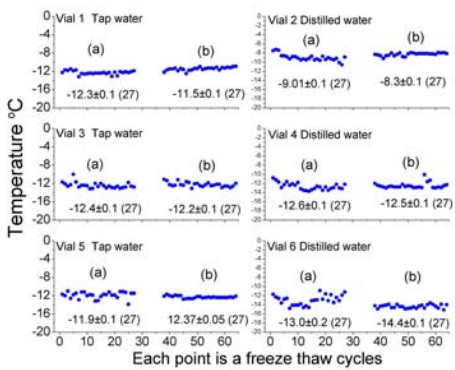

**Fig. 17.** The spontaneous freezing temperatures of the six vials depicted in Fig. 16. In each panel, (a) denotes the data points corresponding to the first 27 freeze cycles, and (b) denotes the data from the 27 freeze cycles performed after the vials were vigorously shaken for five minutes. The numbers under each set of data points are the mean temperatures at which freezing occurred and the standard errors.

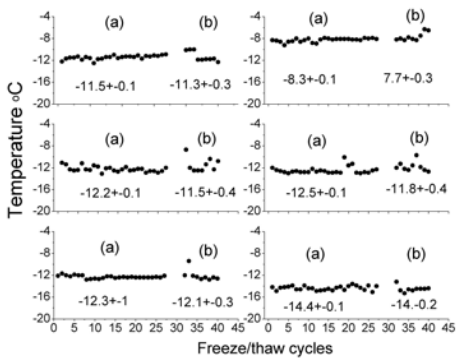

**Fig. 18.** The spontaneous freezing temperatures of the six vials depicted in Fig. 16. In each panel, (a) denotes the data from the 27 freeze cycles prior to turning the vials upside-down, and (b) denotes the data from the freeze cycles after the vials were turned upside-down. The numbers under each set of measurements are the mean temperatures and the standard errors.

Next, the vials were turned upside-down so that the water was in contact with the flame-sealed ends of the vials. Again, no major changes in the spontaneous freezing temperatures were observed; see Fig. 18. Whatever is causing these water specimens to consistently freeze at the same temperature is unique to the water and is not the effect of a particular site in the glass vial.

Next, the vials were removed from the holder and placed in a boiling water bath for two hours. After 22 freeze/thaw cycles, they were heated again for three hours in boiling water, to determine whether heating affects the spontaneous freezing temperature. These results are presented in Fig. 19. Most of the changes in the spontaneous freezing temperature are considered **major changes**. We believe that they are the result of the destruction and/or production of nucleation sites (motes) in the water, as proposed by Dorsey.[6]

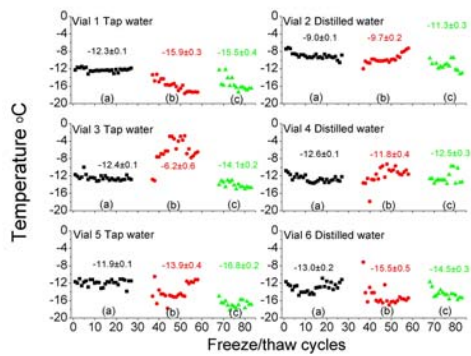

**Fig. 19.** The data points under (a) were collected prior to heating the water to ~100°C, those denoted by (b) were collected after heating and those under (c) were obtained after a second heating. The numbers above each set of points are the mean spontaneous freezing temperatures and the standard errors.

**X. Experiment 7** Biological nucleation agents

In this experiment, fourteen screw-cap vials were placed into a plastic vial holder similar to the one described in Experiment 6, providing the ability to change several variables in addition to the temperature. Snow water, tap water, deionized water and distilled water were used. Some vials were heated, while some were left unheated, as controls. Water was removed from vials and returned to the same vial or a different one. A few drops of snow water were added to distilled water, and distilled water was added to snow water and tap water. The objective of all of this is to determine whether these changes affect the mean spontaneous freezing temperature of the specimen and, if so, how. Some of the results are best summarized in Fig. 20.

To produce the data shown in Fig. 20, 10 ml samples of fresh snow water were added to six vials. Vials 2, 3 and 6 were controls and were not opened or heated for the first five experiments. Each experiment consisted of nine or more freeze/thaw cycles over several days. The objectives: Can the mean spontaneous freezing temperatures be changed, by heating, to new, stable spontaneous freezing temperatures, and can they then be restored by adding nucleation agents from the original stock of snow water?

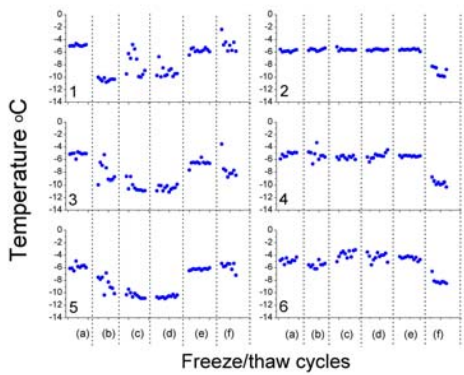

**Fig. 20.** Spontaneous freezing temperatures for six vials containing 10 ml of fresh snow water. After the first set of freezer/thaw cycles (a), vials 1, 3 and 5 were heated to ~100°C three times prior to the dotted lines between (a), (b), (c) and (d). Vials 2, 3 and 6 were controls and were not disturbed until (f).

It is well known that biological agents are involved in the freezing of water in the atmosphere; snow and hail are examples, as is frost on plant tissue.[11] We collected fresh snow and let it melt at room temperature to obtain known biological nucleation agents. The intent of this experiment was to test the hypothesis that the spontaneous freezing temperature can be changed at will. All six vials were initially put through nine freeze/thaw cycles, as in (a) of Fig. 20. Between cycles (a) and (b), vials 1, 3 and 5 were placed in a boiling water bath for one hour and then allowed to return to room temperature, after which all six vials were cycled nine more times. This process was repeated twice between (c) and (d). Between (d) and (e), vials 1, 3 and 5 were opened. 0.5 ml of water was removed and 0.5 ml of snow water from the original stock was added. Between (e) and (f), vials 1, 3 and 5 were opened and the water was removed and returned to the same vial. At this time, the control vials were disturbed for the first time. The snow water was removed, and 10 ml of deionized water was added. The nucleation agent was clearly in the water and was associated with the vial. **This result shows that the spontaneous freezing temperature of water can be changed by heat and then restored by the new biological ice nucleation agents added to vials 1, 2 and 5.**

**XI. Experiment 8** Hot water consistently freezes faster than cool water

Using information gained from Experiments 6 and 7, we were able to create conditions under which hot water froze before cooler water 28 times in 28 attempts. Although we elected to end this run after 28 freeze thaw/cycles over nine days, we believe that we could have continued for much longer. To achieve hot water freezing faster than cool water 28 out of 28 attempts we selected vials 2 and 5 from Fig. 19 because of the ~5.5° C difference in their mean spontaneous freezing temperatures.

If two water specimens have a relatively large difference in spontaneous freezing temperatures, and the one with the highest spontaneous freezing temperature is the warmest when cooling begins, it will more often than not freeze first. Newton's Law of Cooling can be used to calculate the necessary minimum difference in spontaneous freezing temperatures as a function of the two initial temperatures between ~100° C and ~0° C. The volume of water must be small relative to the volume of the cooling environment; in our case, the ratio was 1200 to 1. If the volume of the hot water is large enough to notice the affect of the cooling environment, then the conditions have changed and the experiment is no longer valid.

Immediately after heating the vial 5 to ~100° C in boiling water while vial 2 remained at 25° C or lower, both samples were placed in the freezer to freeze. This process was repeated 28 consecutive times over a nine-day period. We show two typical cooling curves in Fig. 21. Notice that the latent heat of freezing was released from the hot water first, at a higher temperature than the cold water.

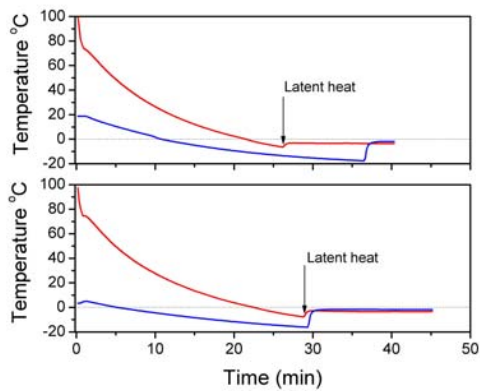

**Fig. 21.** Cooling curves of 10 ml hot and cold water samples flame-sealed in Pyrex test tubes. The onset of the release of the latent heat of fusion is the signal that the water is frozen according to the definition of freezing used in this paper.

These results suggest that it should be possible to observe ~100° C water freezing before ~0° C water. Fresh snow water and deionized water were used in this experiment. Snow water was selected because, in previous experiments, it usually spontaneously froze at higher temperatures than deionized water. Ten millimeters of each water specimen were flame-sealed in six Pyrex test tubes, three of which contained snow water and three of which contained deionized water. The vials were heat-cycled to 100° C several times before the data presented in Fig. 22 was collected. Prior to cooling the vials in the freezer, the three snow water vials sat in boiling water until their temperatures reached ~100° C, while the three deionized water vials sat in a salt/water ice bath. Notice that two of the hot water vials froze before any of the cold water vials. The temperature probes did not read 0° C when the latent heat of freezing was released because they were located on the outside of the vials, and there was just enough latent heat to raise the water temperature to ~0° C, while the vial remained cooler.

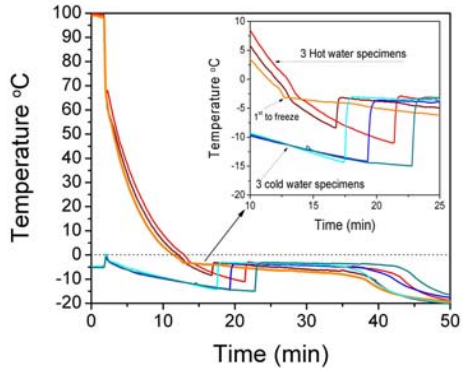

**Fig. 22.** Fresh melted snow water at ~100° C and deionized water at ~0° C were placed in a freezer at time T = 0. Two of the three hot water samples froze before any of the cold water samples froze.

**XII. Experiment 9** Water remains liquid just above its spontaneous freezing temperature.

Four small Pyrex test tubes, each containing two milliliters of tap water, were flame-sealed. The vials were repeatedly heated to ~100° C and then cooled until the water in each vial froze. The process was repeated until the spontaneous freezing temperatures varied by less than 1° C over five or more consecutive freeze/ heat/freeze cycles. The four vials were then placed in a constant temperature bath at -15 °C, ~1° C above the mean spontaneous freezing temperature of the vial that exhibited the highest spontaneous freezing temperature. The purpose of this experiment was to test how long undisturbed tap water will remain liquid at -15° C. The bath's stability is assumed to be ±0.5° C. We believe that it is the spontaneous freezing temperature, not the initial temperature that determines when a specimen of water freezes. If we are correct, then the water in these tubes should remain liquid until their temperatures fall below their previously determined mean spontaneous freezing temperatures. Vial 3, which was the vial whose spontaneous freezing temperature was closest to -15° C, froze on Day 72. The other three samples were still liquid after 167 days. Dorsey[6] held water between -8 and -10° C for over 300 days.[6] The spontaneous freezing temperature of his specimen was -13° C.

**XIII. Conclusions:**

The Mpemba effect, describing the phenomenon of initially hot water freezing before cooler water, occurs only when the water supercools and the cooler water has a lower nucleation temperature than the warmer water. The Mpemba effect is not observed if the hot water freezes first but all conditions during cooling are not "identical," this is the result of increased thermal conductivity. In this case, the hot water container affected the experimental conditions.

Under normal conditions, ice that is warmed from less than 0° C will always begin melting when its temperature reaches 0° C. However, when liquid water is cooled from above 0° C, it often will not begin freezing until it has supercooled to several degrees below 0° C. This is why hot water can freeze before cooler water when all experimental conditions are identical except for the initial temperatures of the water. **Hot water will freeze before cooler water only when the cooler water supercools, and then, only if the** **nucleation temperature of the cooler water is several degrees lower than that of the hot water. Heating water may lower, raise or not change the spontaneous freezing temperature.**

The key is creating **identical** conditions; if the hot water causes better thermal conductivity between the hot water container and the cooling environment, then the conditions are no longer identical. An observed example of this was the melting of freezer frost, shown in Figs. 9 and 10. When conditions for the two samples are identical, the hot water always takes longer to reach $0^\circ$ C, as shown in Fig. 21.

## ACKNOWLEDGMENTS:

The author thanks William R. Gorman for conducting many early exploratory experiments and for fruitful discussions. Thanks also to I. Brownridge, S. Shafroth, B. White, H. Roberson and J. I. Katz for many fruitful discussions and suggestions. This work was supported by Binghamton University's Department of Physics, Applied Physics and Astronomy.

Electronic mail:  jdbjdb@binghamton.edu

# Supplemental Information

## Freezing and melting still water

Under normal conditions, ice that is warmed from less than $0^o$ C will always begin melting when its temperature reaches $0^o$ C. However, when liquid water is cooled from above $0^o$ C, it often will not begin freezing until it has supercooled to several degrees below $0^o$ C. This is why hot water can freeze before cooler water when all experimental conditions are identical except for the initial temperatures of the water. Hot water will freeze before cooler water only when the cooler water supercools, and then, only if the nucleation temperature of the cooler water is several degrees lower than that of the hot water.

## Heating water

Heating water may lower, raise or not change the spontaneous freezing temperature. The keys to observing hot water freezing before cold water are supercooling the water and having a significant difference in the spontaneous freezing temperature of the two water specimens.

## Nucleation

Clean water that is setting undisturbed in a freezing environment (freezer) will not necessarily freeze when its temperature falls to $0^o$ C. It will generally supercool to well below $0^o$ C before heterogeneous nucleation initiates freezing. Small volumes of **very pure** water with no "ice nucleation agents" (foreign agents) can be supercooled to ~ $-40^o$ C; at this temperature it is homogeneous nucleation that initiates freezing.

When freezing is initiated by heterogeneous nucleation the water will freeze when its temperature reaches the "ice nucleation temperature" of the foreign agent with the highest "ice nucleation temperature"; see Fig. 1 below.

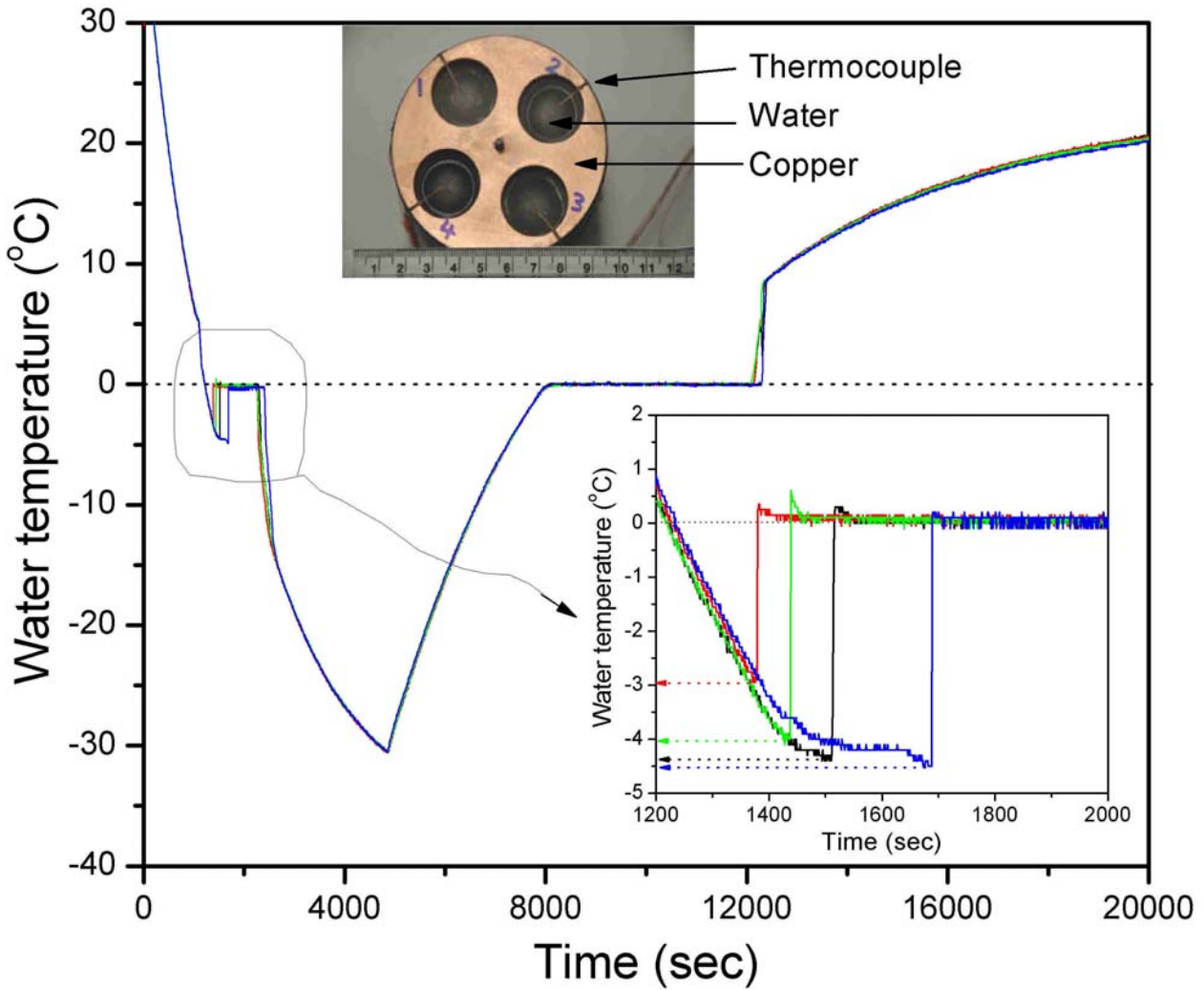

**Fig. 1.** Four specimens of tap hard water in a copper container designed to maintain identical cooling conditions for each specimen.

# Experimental results with known and unknown ice nucleation agents

The known ice nucleation agent is **silver iodide (AgI)**.  Silver iodide is one of the agents used in seed clouding in attempts to increase rain fall.

## A typical glass vial with water

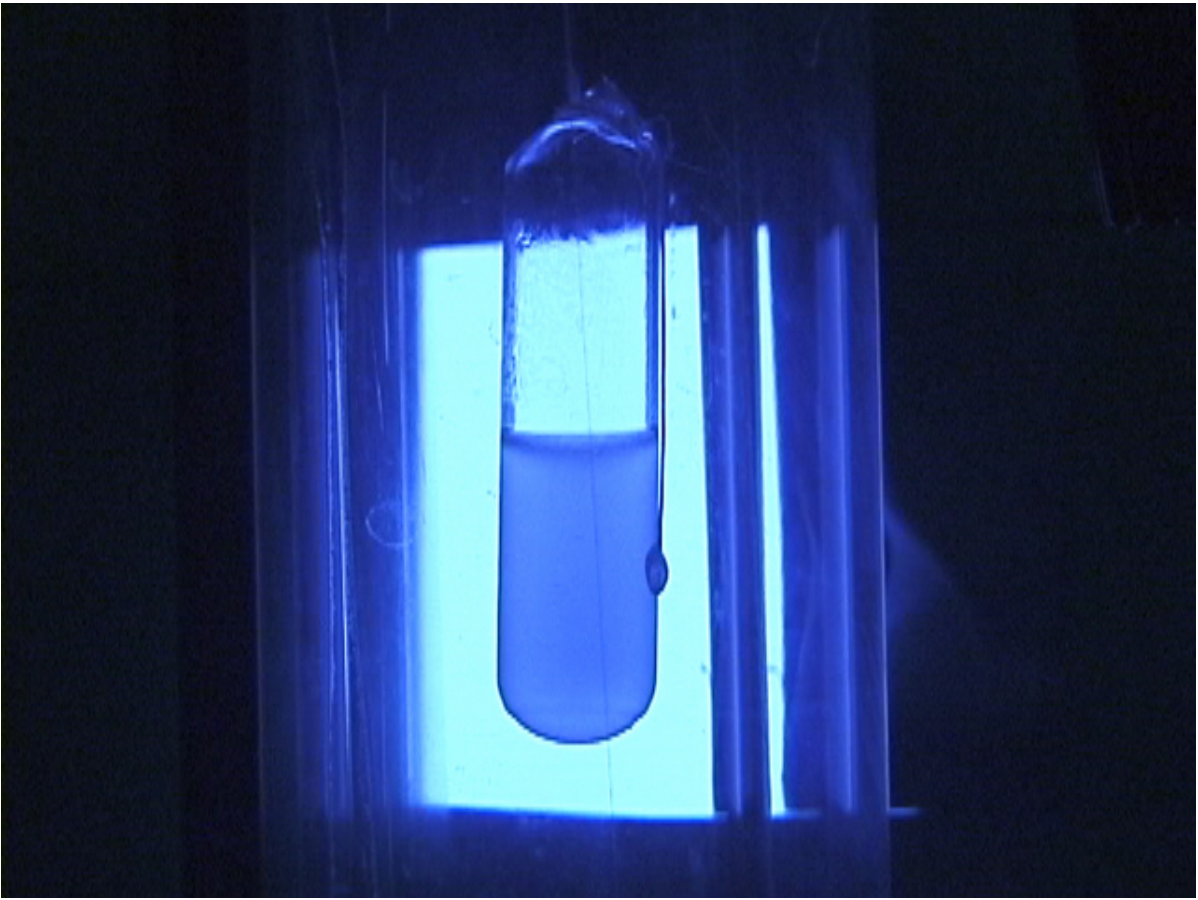

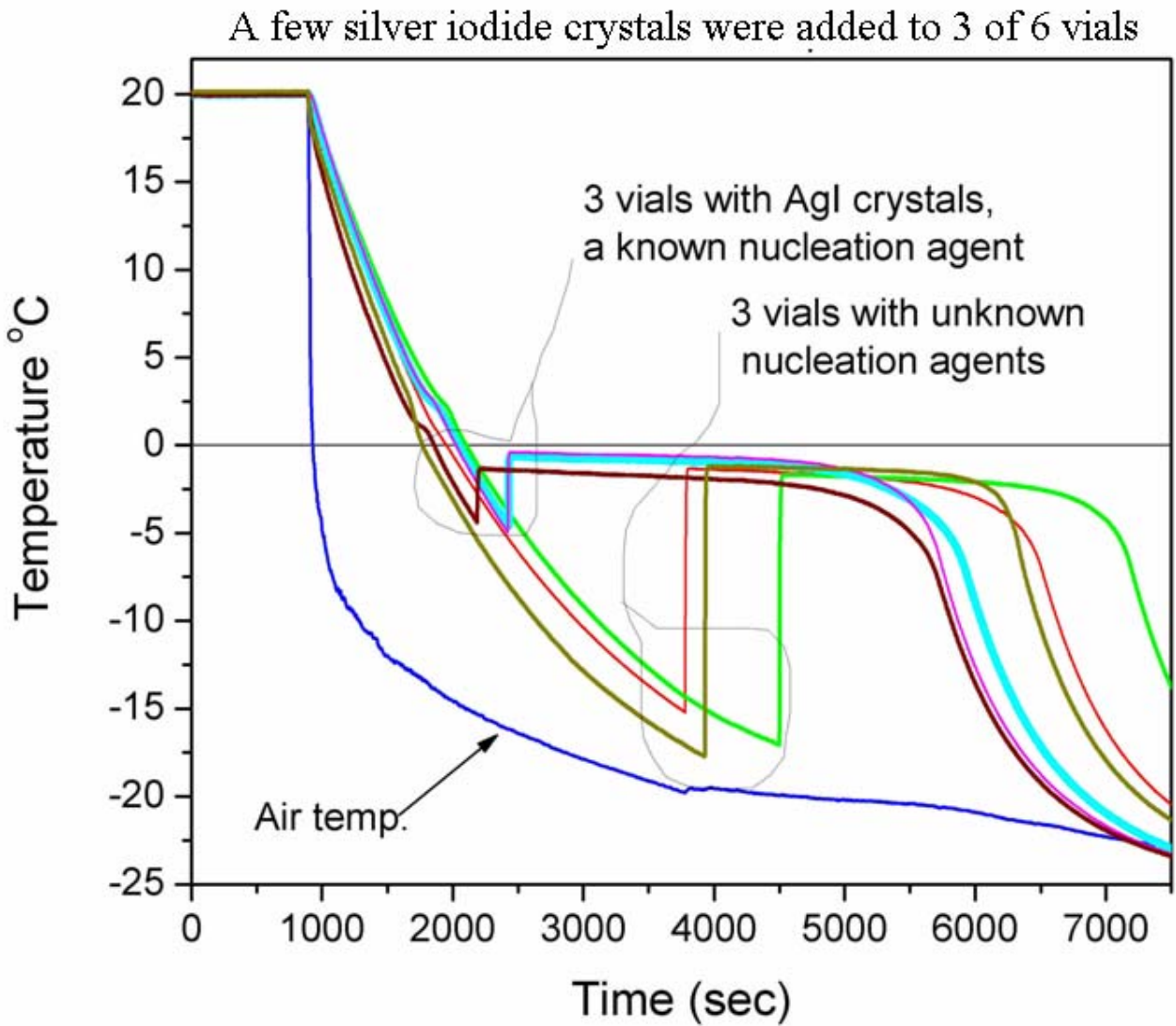

**Fig. 2.** Six vials cooled from 20° C.

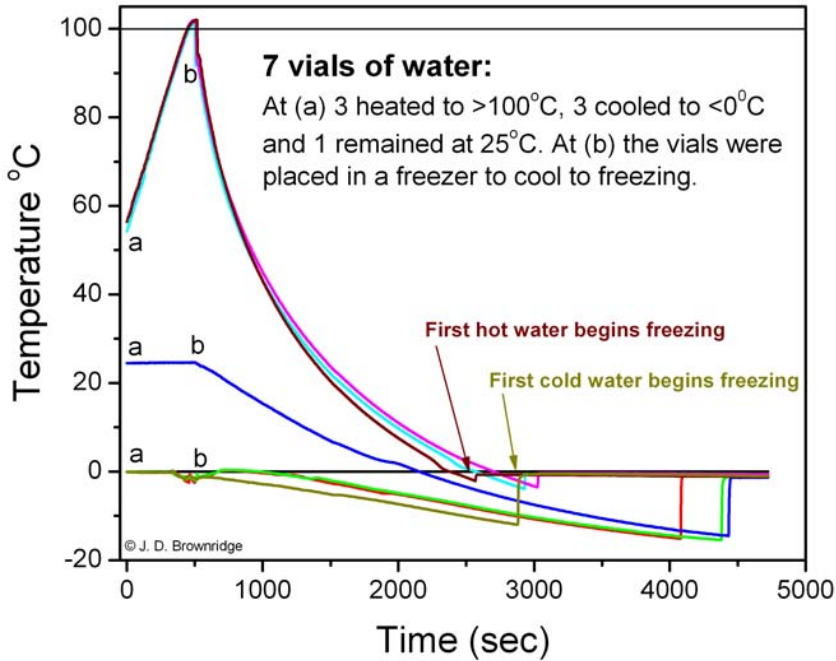
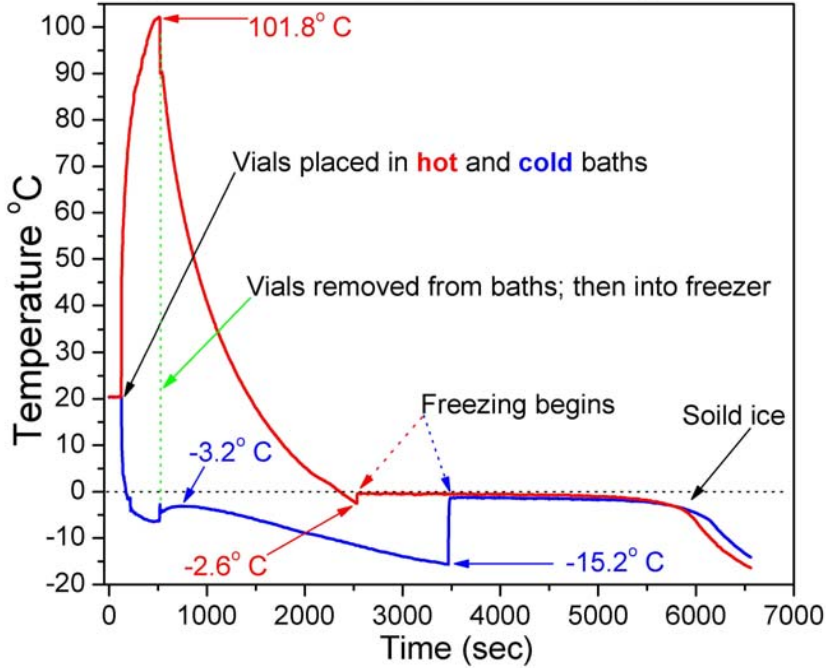

**Figs. 3 and 4.** AgI is the nucleation agents in the hot water. Unknown nucleation agents in the cold water.

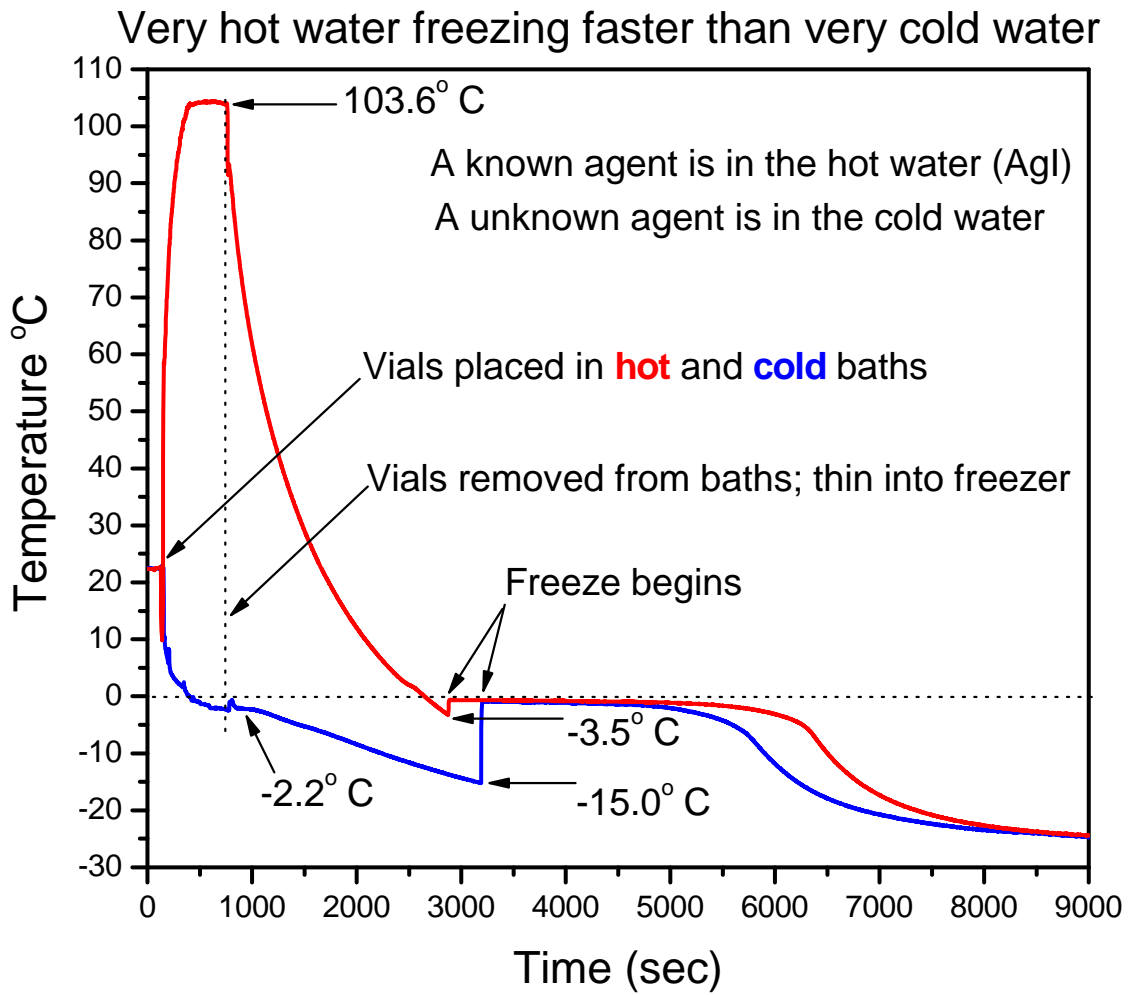

**Fig. 5.** Hot water held above 100° C for several minutes. Cold water held below 0° C for the same length of time.

# Set-up

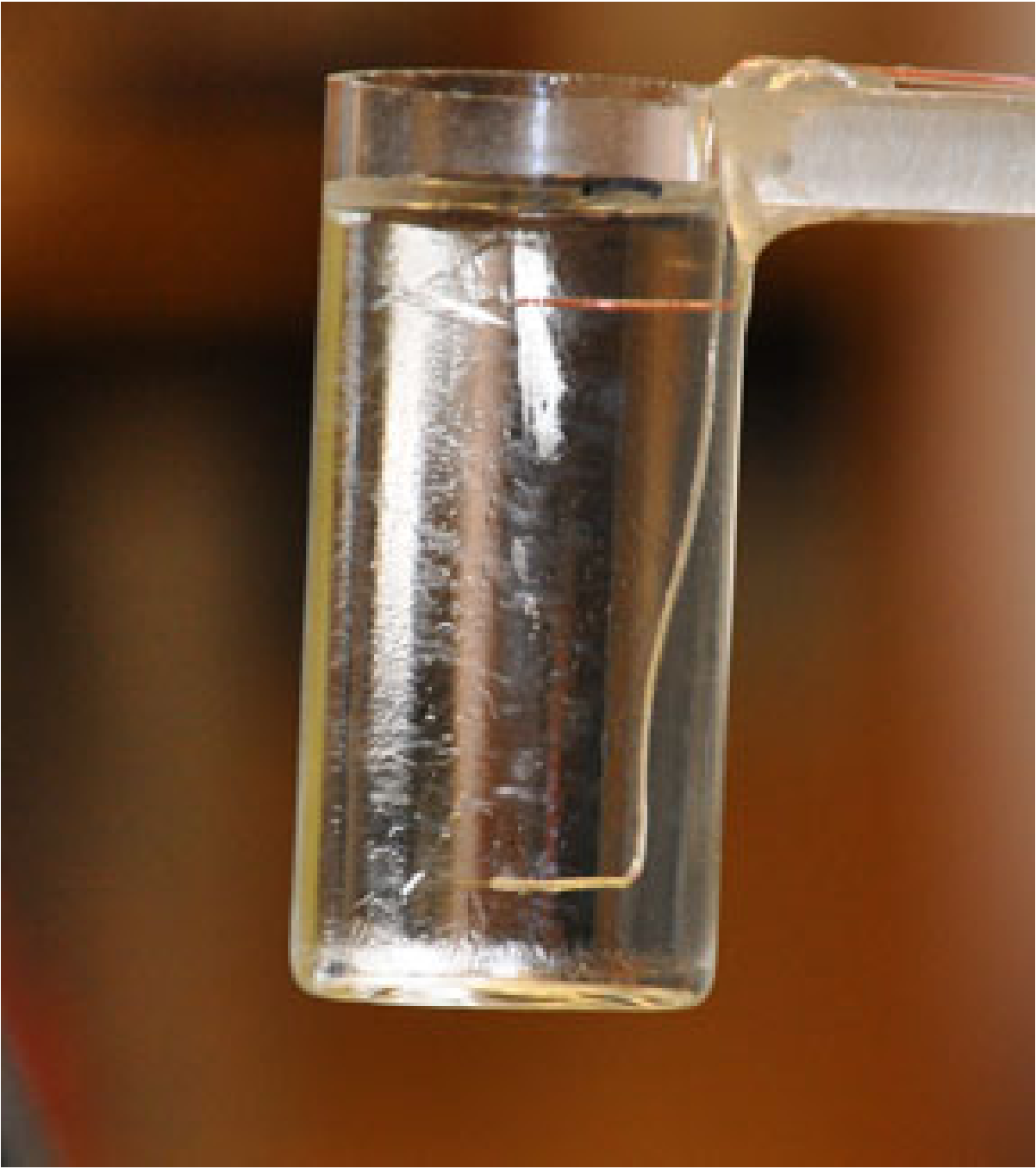

**Fig. 6.** A typical open vial set-up

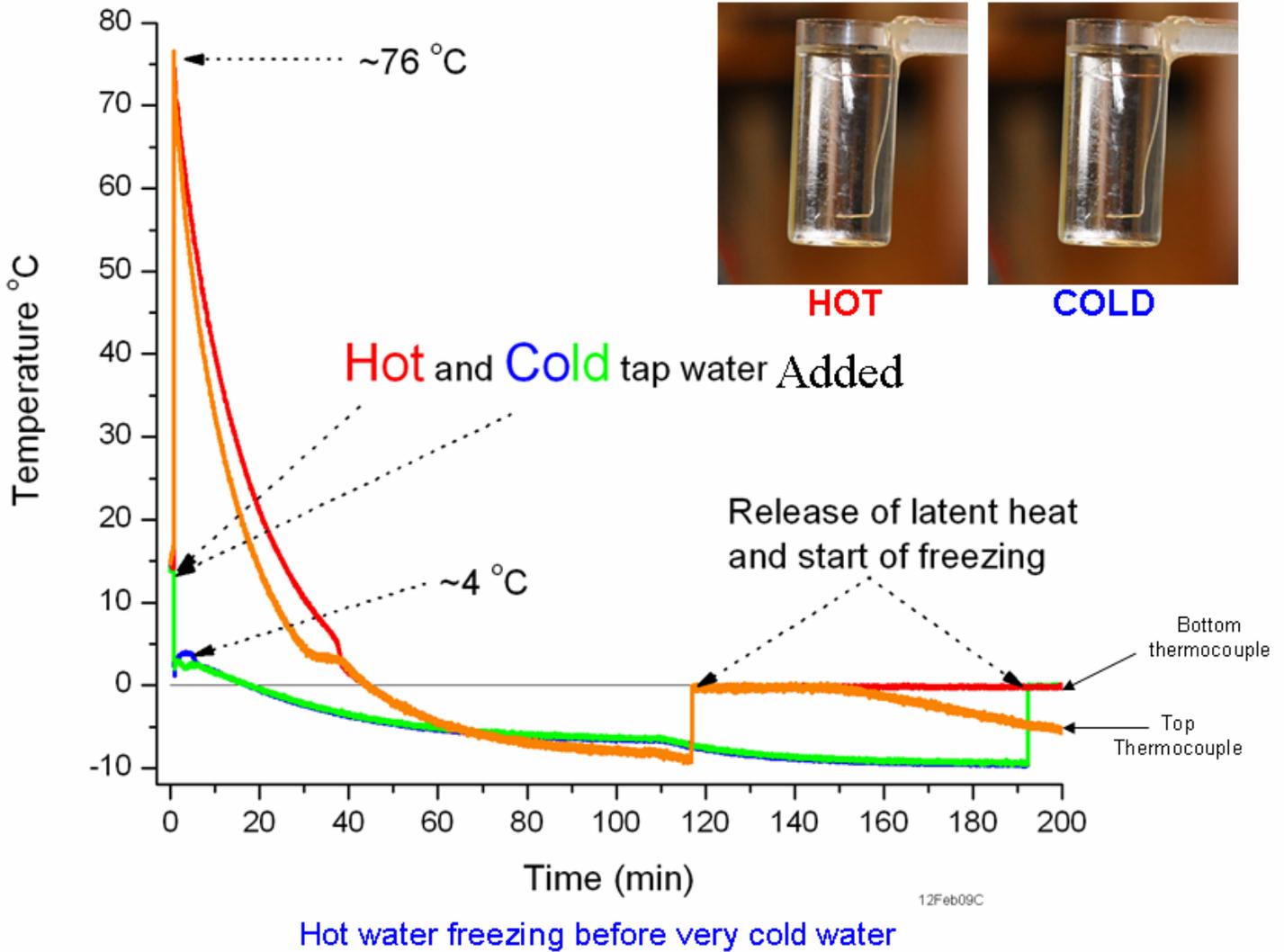

**Fig. 7.** Cooling curves for water at the top and bottom of open vials.

Photos of freezing water (2 ml of water)

The speed of the ice front depends on how low the water is supercooled

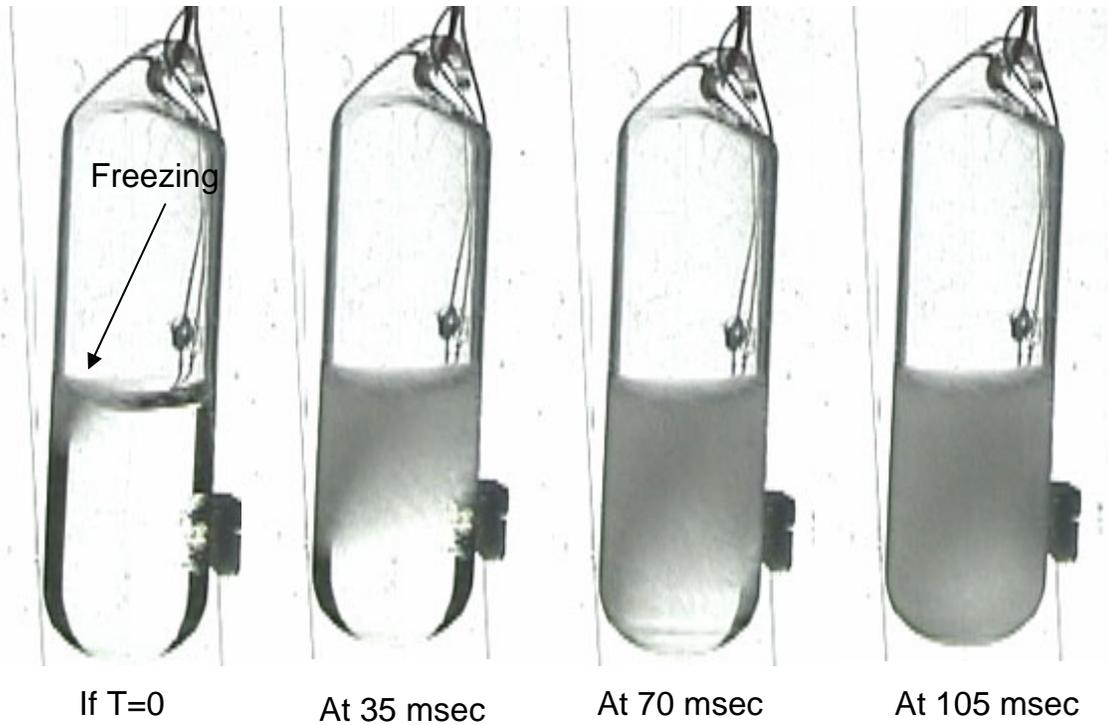

If T=0    At 35 msec    At 70 msec    At 105 msec

**Fig. 8.** Time-lapse photos of water freezing. The deeper the supercooling the faster the freezing front moves.

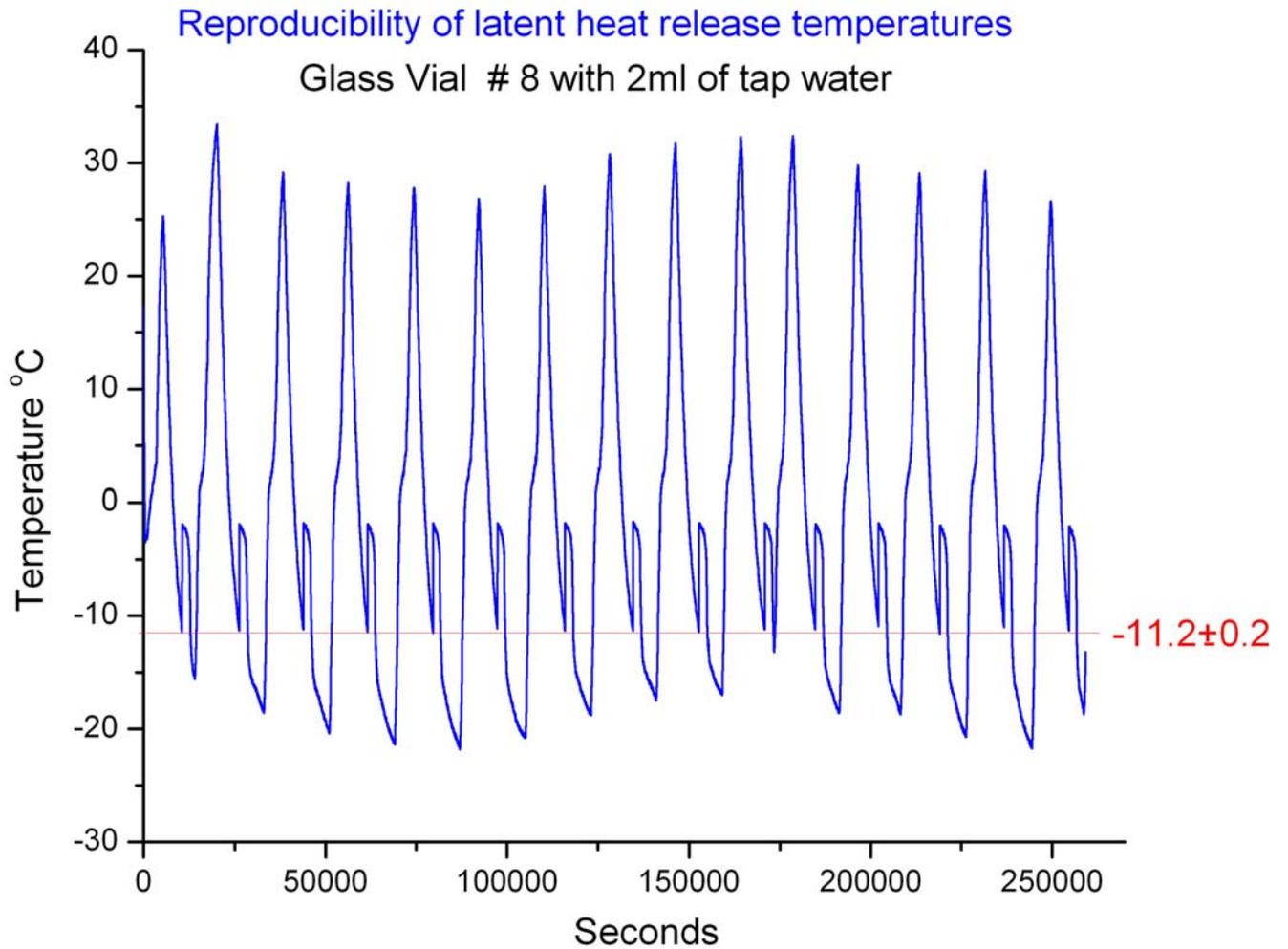

**Fig. 9.** Here water remained undisturbed as it is cycled through 15 freeze/thawed cycles.

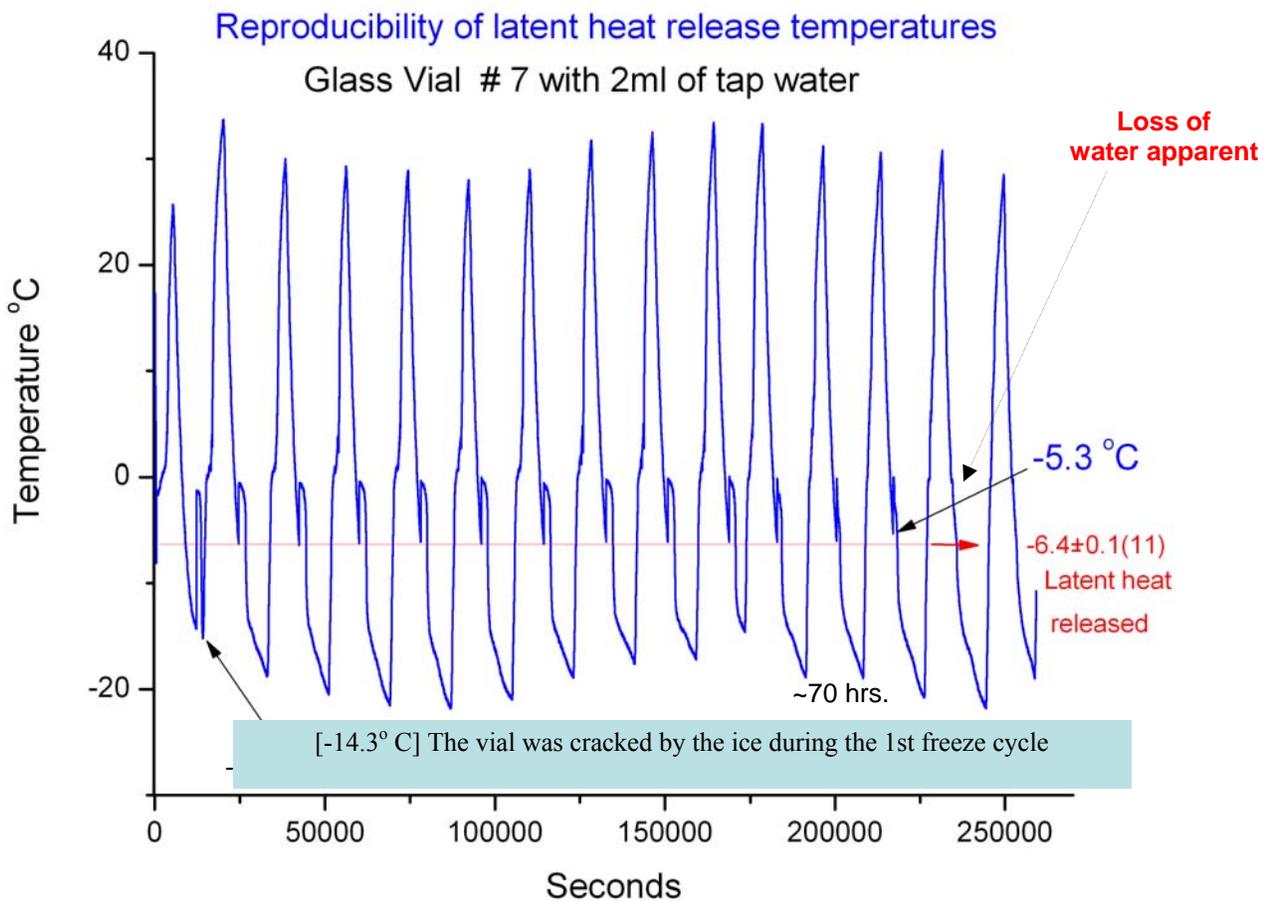

**Fig. 10.** Notice the abrupt change in the latent heat release temperature after the 1st freeze/thaw cycle. A higher temperature "ice nucleation agents" was produced when the glass vial cracked.

**Fig. 11.** Shaking water in a container may change the "spontaneous ice-nucleation temperatures"

| Vial # | Latent heat released at °C | | Net difference |
|---|---|---|---|
| | Initially | After shaking | |
| 1 | -11.0 ± 0.5 | -11.3 ± 0.3 | 0.3 ± 0.6 |
| 2 | -3.3 ± 1.7 | -13.8 ± 0.1 | 10.6 ± 1.7 |
| 3 | -10.5 ± 0.9 | -14.3 ± 0.1 | 3.8 ± 0.9 |
| 4 | -9.1 ± 0.8 | -11.3 ± 0.1 | 2.2 ± 0.8 |
| 5 | -9.9 ± 0.3 | -10.8 ± 0.3 | 0.9 ± 0.4 |
| 6 | -7.5 ± 1.2 | - 9.7 ± 0.1 | 2.3 ± 1.2 |
| 7 | -3.4 ± 1.7 | -1.2 ± 1.1 | -2.2 ± 2.0 |
| 8 | -9.8 ± 0.8 | -9.9 ± 0.2 | 0.1 ± 0.8 |
| | 13 cycles | 13 cycles | |

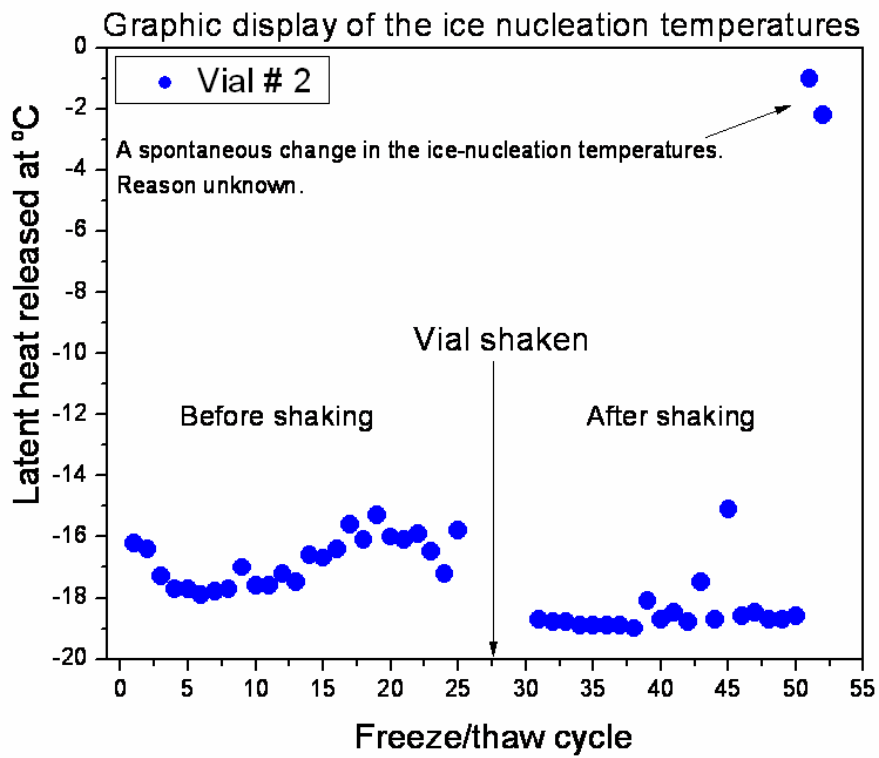

**Fig. 12 and 13.** Graphs of latent heat release temperatures.

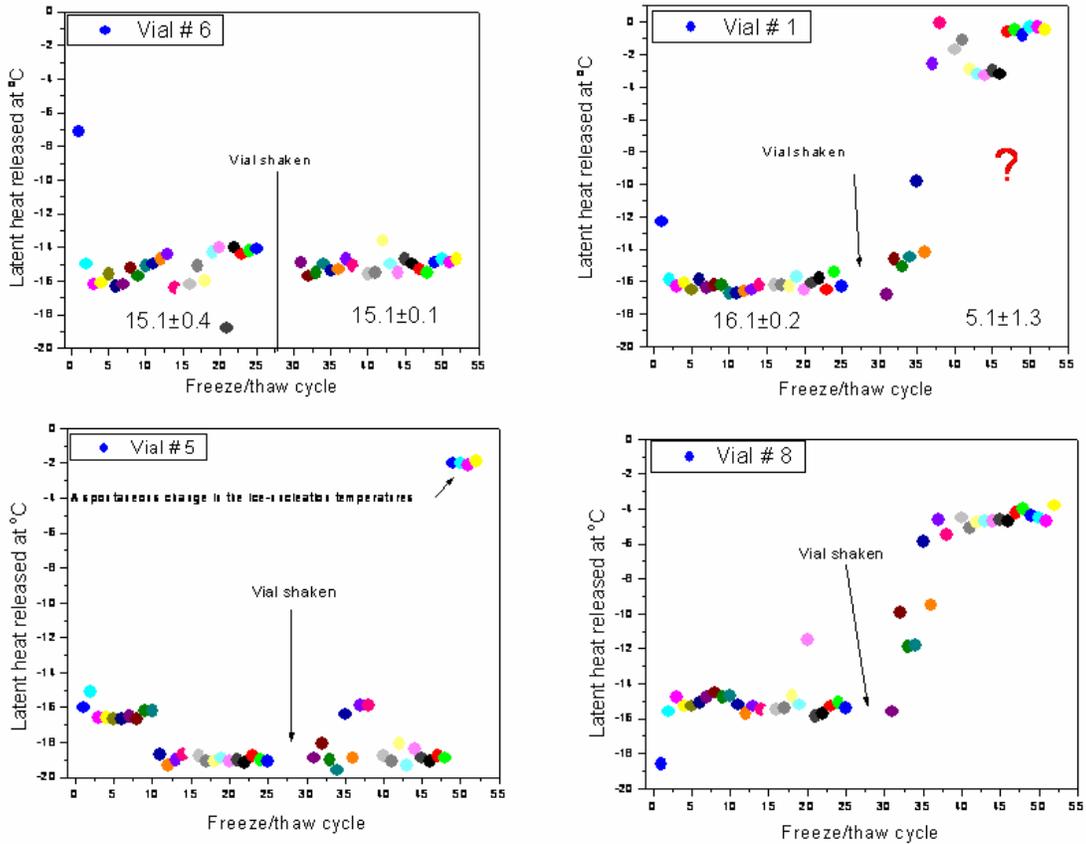

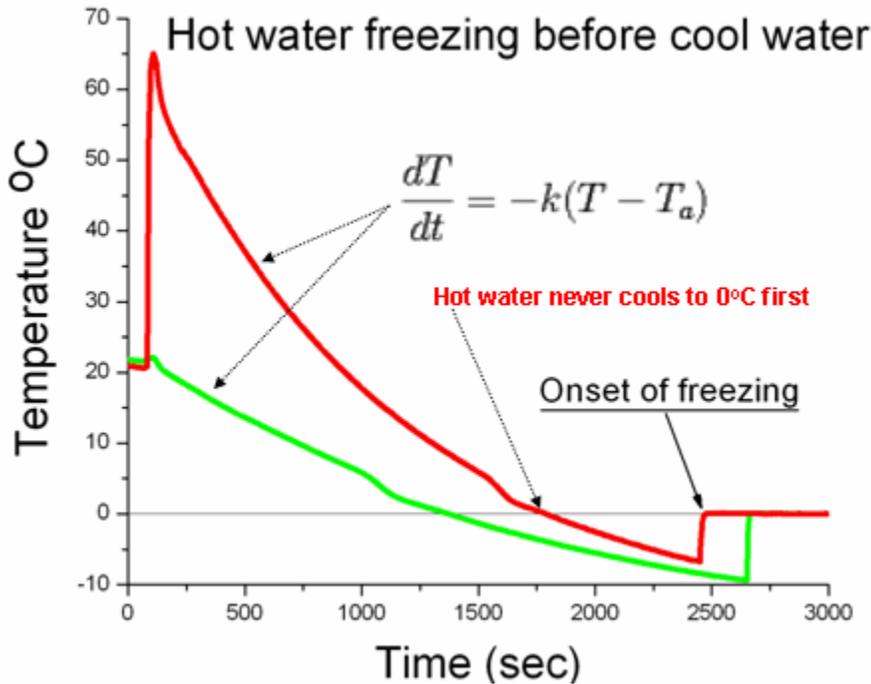

**Fig. 14.** Hot and cool water cooling curves.

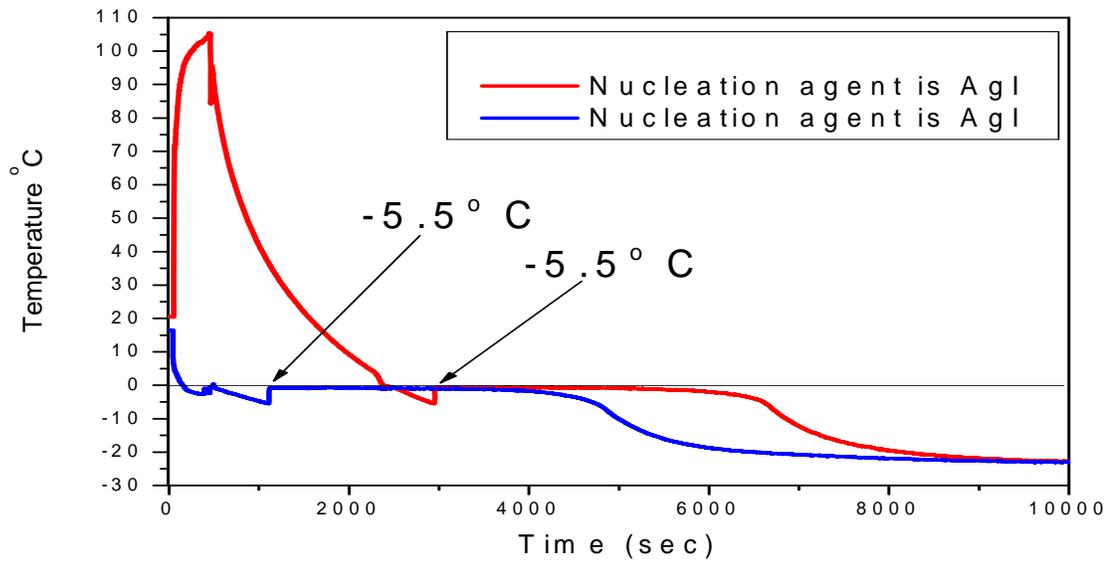

**Fig. 15.**

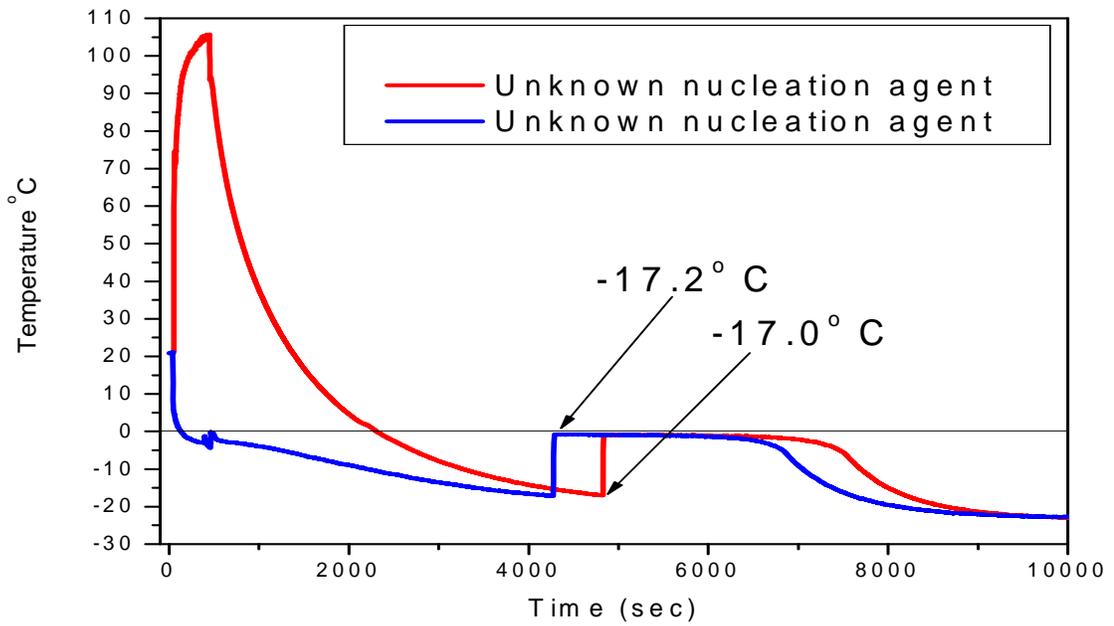

**Fig. 16.**

**Figs. 15 and 16. <span style="color:red">If all other conditions are equal and remain so during cooling; warmer water will not cool to $0^o$ C before cooler water.</span>** "Newton's Law of Cooling"; see Fig. 13. If the waters begins freezing at $0^o$ C then the cooler water will **always** freeze first because its temperature reaches $0^o$ C first.

If both waters supercool, then the water that arrives at the temperature of the "ice nucleation agent" with the highest temperature in it will freeze first as shown above. That is why the cold water froze first in the above Figures.